\documentclass[prd,aps,floats,twocolumn,nofootinbib]{revtex4-1}
\usepackage{slashed}
\usepackage{mathtools}
\usepackage{amsfonts}
\usepackage{amssymb}
\usepackage{epsfig}

\usepackage{bm}

\begin{document}

\newcommand{\m}[1]{\mathcal{#1}}
\newcommand{\nn}{\nonumber}
\newcommand{\ph}{\phantom}
\newcommand{\eps}{\epsilon}
\newcommand{\be}{\begin{equation}}
\newcommand{\ee}{\end{equation}}
\newcommand{\bea}{\begin{eqnarray}}
\newcommand{\eea}{\end{eqnarray}}
\newtheorem{conj}{Conjecture}

\newcommand{\plk}{\mathfrak{h}}


\title{Mach's principle and dark matter}

\date{}

\author{Jo\~{a}o Magueijo}
\email{j.magueijo@imperial.ac.uk}
\affiliation{Theoretical Physics Group, The Blackett Laboratory, Imperial College, Prince Consort Rd., London, SW7 2BZ, United Kingdom}

\begin{abstract}
In this paper we entertain a Machian setting where local physics is non-locally affected by the whole Universe, taking the liberty to identify the local (``Newton's  bucket'') with our visible Universe, and the whole Universe (Mach's ``fixed stars'') with the global Universe beyond our horizon.
Crucially, we allow for the two to have different properties, so that we are beyond the traditional FRW setting. For definiteness we focus on theories where non-locality arises from evolution in the laws of physics in terms of spatially global time variables dual to the constants of Nature. Since non-local theories are foliation-dependent, the {\it local} (but not the global) Hamiltonian constraint is lost. This is true not only while non-locality is taking place, but also after it ceases: the local Hamiltonian constraint is only recovered up to a constant in time, keeping a memory of the integrated past non-locality. We show that this integration constant is equivalent to preserving the local Hamiltonian constraint and adding an extra fluid with the same cosmological properties as conventional pressureless dark matter. The  equivalence breaks down in terms of clustering properties, with the new component attracting other matter, but not budging from its location. This is the ultimate ``painted-on'' dark matter, attracting but not being attracted, and nailing down a preferred frame.


\end{abstract}

\maketitle

\section{Introduction}
In this paper we seek to relate the enigma of dark matter and an issue at the core of the foundations of physics:  the awkward interaction between General Relativity (GR) in its final form and the Machian principle that provided some of its inspiration~\cite{MachBook,MachReview}. In trying to extend relativity of motion to non-inertial frames, Mach implied that Newton's rotating bucket would not feel inertial forces if a presiding Universe were not present. 
But by making inertial forces the result of  relative acceleration between the local and the global Universe, the door opened up to ``pantheistic'' direct interactions between the whole Universe and its parts. Ironically this introduces a preferred frame, albeit one determined by matter (the ``fixed stars''), as well as non-local interactions (action at a distance with infinite speed). Hence Mach's views, when interpreted in this fashion (see~\cite{MachReview} for alternatives), are in contradiction with the GR it inspired, within which the  metric field is part of reality and suffices to {\it locally} explain the inertial forces. In this paper we revisit the possibility of global actions on local physics, thereby ``returning to Mach''  in a sense.

Mach never elevated his principle to a  theory (although the interactions he envisaged were manifestly gravitational). What follows is decoupled from the issue of inertial forces, but preserves the idea of a theory with global interactions.  Furthermore, although most of this paper applies to any theory where there are global variables interacting with local physics, for concreteness we focus on theories where the global variables are canonical pairs made up of the ``constants'' of Nature
and their dual clocks~\cite{JoaoPaper}. A prototype for this construction is the Henneaux-Teitelboim formulation~\cite{unimod}  of unimodular gravity~\cite{unimod1,UnimodLee1,alan,daughton,sorkin1,sorkin2}, where the cosmological constant $\Lambda$ is the target constant to be demoted to an on-shell constant, with a canonical dual which on-shell is the 4-volume to the past, or unimodular time~\cite{Bombelli,UnimodLee2,JoaoLetter,JoaoPaper}. Other targets can be chosen~\cite{pad,pad1,vikman,vikman1,vikaxion,JoaoPaper,evolution}. Usually the ``times'' ticked by these clocks do not appear  in the action, presenting as Lagrange multipliers enforcing the on-shell constancy of the constants. But we can envisage {\it evolution} in the laws of physics in terms of such time variables~\cite{evolution}, leading to concomitant variations in the constants of Nature to which they are tied. As in Mach's vision, in such theories there are global cosmic variables which interfere with the local physics, a situation we have explored in the context of black hole formation in~\cite{BHevolution}. 

All such non-local theories peg down a global foliation of space-time into space and time. 
This implies that the lapse function must be a spatially global variable, so as to preserve the foliation. As a result there is fundamentally no local Hamiltonian constraint, but only a global Hamiltonian constraint.
Furthermore, 
as proved in Sections~\ref{locglob}--\ref{ex}, whenever global interactions are switched off, the local Hamiltonian constraint is not restored, but, rather, one obtains the weaker result that the local Hamiltonian is conserved. Hence, it becomes a generally non-vanishing constant in time (which can vary in space),  representing the integrated memory of all past evolution or any other global interactions. We can absorb this constant into a redefined Hamiltonian satisfying a Hamiltonian constraint, containing a new matter component, which we will call {\it Machian matter}. The fact that such matter mimics some, but not all aspects of cold dark matter, is the point of this paper. Specifically it reproduces all the gravitational effects of dark matter as a homogeneous and isotropic background component (Sections~\ref{ex} and~\ref{illustration}). In terms of inhomogeneities, however, it is strikingly different (Sections~\ref{contrast} and~\ref{concs}), providing grounds for testability and phenomenology.

\section{Local and global variables and the Hamiltonian constraint}\label{locglob}
Although what follows distils lessons learnt in~\cite{BHevolution}, we start by presenting the argument in general, abstracting from concrete theories and from the Machian setting. We consider a set of regions indexed by $I$, with coordinate volume $V_I$ and $V_\infty=\sum_I V_I$. In general these regions could be infinitesimal volumes around any point (with 
$V_I\rightarrow d^3x$ and the summation replaced by an integral) but in simplified models $I$ indexes regions where a symmetry reduction has taken place. For example, in~\cite{BHevolution} it indexes an inner collapsing FRW model, a Schwarzchild bubble and an outer FRW Universe (see also~\cite{OS}).
We consider $V_I$ to be {\it coordinate} volumes because the Lagrangian is assumed to be a density, so that with this definition the action is a scalar. The $V_I$ are also then constant.

We define global variables $\{\alpha, T_\alpha\}$, as opposed to local variables $\{q_I,p_I\}$, from the structure of the action:
\begin{eqnarray}
     S&=&\int dt\,  [ V_\infty \dot\alpha T_\alpha\nn\\
     &&+ \sum_I V_I(\dot q_I p_I-N H_I(q_I,p_I;\alpha,T_\alpha)) +...] .\label{split}
\end{eqnarray}
Here $N$ and $H_I$ are the lapse function and Hamiltonian, and the ellipse denotes other constraints. A foliation is implicit in this 3+1 splitting. In this foliation the global variables do not depend on $I$, i.e. they are spatially uniform. This can be a generic assumption for global variables, but as we will see in Section~\ref{ex-theories}, 
in unimodular-like theories it follows from solving for the Lagrange multiplier ${\cal T}^i$, to eliminate the gauge-dependent modes, with gauge-invariance present to enforce no new local degrees of freedom. The volume factors appear because the variables are defined as {\it intensive}  densities (as is norm for local variables). As a result, the non-vanishing Poisson brackets carry the respective volume factors:
\begin{eqnarray}
    \{q_I,p_I\}&=&\frac{1}{V_I}\\
    \{\alpha,T_\alpha\}&=&\frac{1}{V_\infty}.
\end{eqnarray}
The volume factors are usually neglected (except upon quantization~\cite{hbarrow,Afshordi,realkod,HHpacks,Vilpacks,laurent}) because in local theories they cancel out in Hamilton's equations, as they do here in the equations for $q_I$ and $p_i$. However, they do matter for the global variables' equations of motion:
\begin{eqnarray}
    \dot\alpha&=&N\sum_I \frac{V_I}{V_\infty}\frac{\partial H_I}{\partial T_\alpha}\label{alphaeq}\\
    \dot T_\alpha&=&-N\sum_I \frac{V_I}{V_\infty} \frac{\partial H_I}{\partial \alpha}.\label{Talphaeq}
\end{eqnarray}
As we see, the larger the $V_I$ the more the region $I$ contributes to the dynamics of the global variables. 
We can define a global Hamiltonian {\it density}:
\begin{equation}\label{Htot}
     H=\frac{1}{V_\infty}\sum _I V_I H_I,
\end{equation}
to repackage these equations as 
\begin{eqnarray}
    \dot\alpha&=&N\frac{\partial H}{\partial T_\alpha}\label{ham1tot}\\
    \dot T_\alpha&=&-N\frac{\partial H}{\partial \alpha}.\label{ham2tot}
\end{eqnarray}
Usually the lapse function $N$ can depend on $I$, but given our definition of global variables and their appearance in the local action, $N$ should be foliation preserving, that is $N_I(t)\equiv N(t)$. This is a general feature of theories which break foliation invariance (Horava-Lifschitz theory providing an example~\cite{HL}).
Hence, we lose the local Hamiltonian constraint; by varying with respect to $N$ we still find a global Hamiltonian
constraint:
\begin{equation}
    H=0,
    \label{hamconstr}
\end{equation}
however, no $H_I=0$ constraint exists a priori. We can diagnose the loss of the local Hamiltonian constraint through the loss of the secondary constraint it would have generated.
Since $H_I=H_I(q_I,p_I;\alpha,T_\alpha)$ we have:
\begin{eqnarray}
   \dot {H}_I &=&
   \frac{\partial {H}_I}{\partial q_I}\dot q_I +\frac{\partial {H}_I}{\partial p_I}\dot p_I+ 
   \nn
   \frac{\partial {H}_I}{\partial{\bm \alpha}}\dot {\bm \alpha}+
   \frac{\partial {H}_I}{\partial{\bm T}_{\bm \alpha}}\dot {\bm T}_{\bm \alpha}\nn\\
   &=&\frac{\partial {H}_I}{\partial{\bm \alpha}}\frac{\partial H}{\partial {\bm T}_{\bm \alpha}}-
   \frac{\partial {H}_I}{\partial{\bm T}_{\bm \alpha}}\frac{\partial H}{\partial {\bm \alpha}},
   \label{dotcalH}
\end{eqnarray}
where we have used Hamilton's equations for the local and global variables. The first two terms cancel, but in general the last two do not\footnote{The obvious exceptions are the case where $\dot\alpha=0$ and $H_I$ does not depend on $T_\alpha$, or else the case where there is only one $I$, such as the FRW approximation studied in~\cite{evolution}.}, so:
\begin{equation}
    \dot {H}_I\neq 0
\end{equation}
invalidating the possibility of a local Hamiltonian constraint. Summing over $I$, we still find:
\begin{equation}
    \dot H=\frac{1}{V_\infty}\sum_I V_I \dot H_I=0
\end{equation}
as it should, since $H=0$.

The central question now is, do we recover the local Hamiltonian constraint (and full diffeomorphism invariance) once global interactions switch off?
For example, if the global variables are the constants of nature and their conjugate times, the dependence of $H_I$ on such times may be confined to restricted epochs (say $t_i<t<t_f$). What happens for $t>t_f$, once Machian interactions switch off?  The answer is that, if there ever were Machian interactions (e.g. evolution), the local Hamiltonian constraint is never fully recovered. Eq.~(\ref{dotcalH}) implies $\dot H_I=0$ when $H_I$ does not depend on at least one of the variables in each global canonical pair, but this integrates to:
\begin{eqnarray}\label{DeltaH}
     H_I (t)&\equiv&-m_I\nn\\
     &=&
     \int^{t_f}_{t_i} dt\, \left(  \frac{\partial {H}_I}{\partial{\bm \alpha}}\dot {\bm \alpha}+
   \frac{\partial {H}_I}{\partial{\bm T}_{\bm \alpha}}\dot {\bm T}_{\bm \alpha}\right)+H_I(t_i)
\end{eqnarray}
for $t>t_i$. An integration constant, therefore,
keeps memory of past interactions.

We can redefine the Hamiltonian as:
\begin{equation}
    H_I\rightarrow \Tilde{H}_I=H_I+m_I =0
\end{equation}
and enforce $\tilde H_I=0$ by promoting the global lapse function to a local one, $N=N_I(t)$. We thus recover the local Hamiltonian constraint and full diffeomorphism invariance. However this is done at the expense of introducing a new term in the action and Hamiltonian, which is intrinsically tied to a preferred frame.  
The fact that this integration constant is akin to some aspects of dark matter is the point to be made in this paper.

\section{The Machian setting}\label{Machset}
The discussion in the previous Section was kept general, but in a Machian setting we can further group the local variables and  Hamiltonian terms labelled by $I$ into those pertaining to the region $L$ we want to study (the Mach/Newton ``bucket'', a collapsing star~\cite{OS,BHevolution}, etc)
and those composing the presiding Universe $U$
(the ``fixed stars'', the outer FRW model in~\cite{BHevolution}, etc). The separation is conceptual and part of the modelization (the Machian Universe being, after all, made up of a large number of coarse-grained local regions), with the only condition being $V_L\ll V_U\approx V_\infty$. This leads to a corresponding split in the local terms in action and Hamiltonian, so that 
the total Hamiltonian density defined in (\ref{Htot}) can be approximated as:
\begin{equation}\label{HUapprox}
    H=\frac{V_U}{V_\infty}H_U+\frac{V_L}{V_\infty}H_L
    \approx H_U
\end{equation}
with $H_U$ and $H_L$ collecting all the $I$ contained in the corresponding sets. This has two important implications. First, (\ref{ham1tot}) and (\ref{ham2tot}) can be approximated by:
\begin{eqnarray}
\dot {\bm \alpha}&\approx& N(t) \frac{\partial { H}_U}{\partial {\bm T}_{\bm \alpha}}\label{GlobalApprox1}
\\
 \dot {\bm T}_{\bm \alpha}   &\approx&-N(t) \frac{\partial { H}_U}{\partial {\bm \alpha}}.\label{GlobalApprox2}
\end{eqnarray}
As pointed out before, the importance of a given local region in determining the global variables is weighted by its volume. Hence, in a Machian setting, the presiding Universe wins by volume and it is to a very good approximation the sole driver of the global variables (which, thus, could have been added to group $U$ for all practical purposes). This implies that in the study of a local region the global variables can be seen as {\it external}, i.e. time dependent variables fixed by a dynamics alien to the local variables under study. 

A second remarkable implication is that, within this approximation, if we insert 
(\ref{HUapprox}) into (\ref{hamconstr}) we find that 
the Machian Universe is subject to an approximate Hamiltonian constraint:
\begin{equation}
    H_U\approx 0.
\end{equation}
The fact that we always have a global Hamiltonian constraint
implies that the presiding Universe is still approximately subject to one too. This is because the ``global'' is effectively identified with the presiding Universe. 

These are general features of theories with global variables in a Machian setting. As explained, the separation between $L$ and $U$ is conceptual. 
In this paper we make a very special choice for $L$. 
We note that unlike Mach's fixed stars, the presiding Universe does not need to be the observable Universe. Indeed, if there is a cosmological horizon effect, it is not. 
By Occam's razor we should assume that whatever lies outside our cosmological horizon, and dominates the global variables' dynamics, has the same average properties we observe in a smoothed version of our horizon, but this is not necessary. 
We toy with this possibility in this paper. We will take our observable FRW patch for local region $L$, ``the bucket'' as it were. By shifting the presiding Universe outside our horizon and allowing it to have a separate dynamics, we are releasing the visible Universe from the Hamiltonian constraint. 

Note that this is necessary for non-trivial local effects. None of the points made in the previous Section are felt in a ``minisuperspace scenario'', or in any scenario with spatial homogeneity built-in. In that case, there is only one region $I$, so that local and global are confused and none of the effects derived are felt (this is also the case for the models investigated in~\cite{evolution}).

\section{Theories, models and scenarios}\label{ex}
Having kept the discussion general 
we now provide examples of theories, models and scenarios fitting into the framework of the last two Sections. 

\subsection{Theories}\label{ex-theories}
For concrete theories with global variables, we choose the ``deconstantization'' procedure in~\cite{JoaoPaper}, itself a generalization of~\cite{unimod}. Hence, for global variables  w$\bm{\alpha}$ we take a set of ``constants'' of Nature. The procedure consists in adding to a base action $S_0$ (which can be GR) a new term:
\begin{eqnarray}\label{S0presc}
      S&=& \int d^4x\, (\partial_\mu {\bm \alpha})\cdot {\cal T}_{\bm\alpha}^\mu +S_0,
\end{eqnarray}
where ${\cal T}_{\bm\alpha}$ is a density. This density is then used~\cite{unimod,Bombelli,sorkin1,sorkin2,UnimodLee2,JoaoLetter,JoaoPaper} to produce a foliation-dependent time variable as follows.
Performing a 3+1 spacetime split, and 
solving for the Lagrange mutiplier ${\cal T}^i$, we first obtain spatial constancy of $\bm \alpha$ on the leaves $\Sigma_t$ defined by the split.  For each entry in the vector $\bm\alpha$, 
the theory has gauge-invariance under  \begin{equation}
    {\cal T}_{\bm\alpha}^\mu\rightarrow {\cal T}^\mu_{\bm\alpha}+\epsilon^\mu_{\bm\alpha}\quad {\rm with}\quad \partial_\mu\epsilon_{\bm\alpha}^\mu =0,
\end{equation}
a feature essential in making sure that the theory has no new local degrees of freedom (see~\cite{unimod} for details), and hence it is purely a theory of global variables.  Retaining only the gauge invariant zero-mode of ${\cal T}^0$ we define a time variable:
\begin{equation}\label{timedef}
    {\bm T}_{\bm\alpha}(t)=\frac{1}{V_\infty}\int_{\Sigma_t} d^3x\, {\cal T}^0_{\bm\alpha}.
\end{equation}
The first term in (\ref{S0presc})  then reduces to the first term in (\ref{split}). As explained before, in the definition (\ref{timedef}) we divide by the spatial coordinate volume $V_\infty$ of each leaf so that time is intensive. For example, in the unimodular case, for which the target $\alpha$ is $\rho_\Lambda$ (the vacuum energy), the time $T_\Lambda$ is 4-volume per unit of coordinate (comoving, in the case of FRW) 3-volume.

For the rest of this paper  we will 
take for $S_0$ the Einstein-Cartan theory with a mixture of perfect fluids (including a cosmological constant) as matter source. This is then adapted to have a lapse $N$ that is purely time dependent. 
As targets for deconstantization  
we choose a selection of:
\begin{eqnarray}
 {\bm \alpha}&=&\left(\rho_\Lambda, \frac{3 c_P^2}{8\pi G} \right),\label{alpha}
\end{eqnarray}
where $\rho_\Lambda$ is the vacuum energy, $c_P$ is the speed of light as it appears in the gravitational commutation relations~\cite{evolution,JoaoPaper,JoaoLetter}, and $G$ is the gravitational constant. This is similar 
to sequestration~\cite{padilla,pad,pad1,lomb} and leads to associated times:
\begin{eqnarray}
T_1(\Sigma_t)\equiv T_\Lambda(\Sigma_t)&=&
  -\frac{1}{V_\infty}\int^{\Sigma_t}_{\Sigma_0} d^4 x \sqrt{-g}\label{TLambda},
\\
    T_2(\Sigma_t)\equiv T_R(\Sigma_t)&=&\frac{1}{6V_\infty}\int^{\Sigma_t}_{\Sigma_0} d^4 x \sqrt{-g} R\label{TRicci},
\end{eqnarray}
i.e.: unimodular and Ricci time. 
Here the 4-volume integration goes from some reference zero-time leaf $\Sigma_0$ to leaf $\Sigma_t$.

It is then important that $S_0$, or equivalently $H$, depend on time(s)
$\bm{T}_{\bm \alpha}$. Otherwise (\ref{alphaeq}) implies that the $\bm{\alpha}$ really are spacetime constants, albeit on-shell only, and any leaf dependence disappears. Properly constant $\bm{\alpha}$ are the quintessential global variables (a matter discussed in ~\cite{padilla,pad1} in terms of fluxes), but that is not the point of this paper, which seeks a Machian interaction with global variables which are only {\it spatially} constant.

We will therefore assume a $\bm{T}_{\bm \alpha}$ dependence in $H$, and further that this happens via constants/parameters $\bm\beta$ other than $\bm{\alpha}$. That the sets $\bm\alpha$ and $\bm\beta$ are disjoint is assumed for simplicity (it avoids introducing a second class constraint). Thus we consider functions $\bm\beta({\bm T}_{\bm\alpha})$, similar to evolution potentials~\cite{evolution}. With these further assumptions, Eq.\eqref{DeltaH} can be written as:
\begin{eqnarray}\label{DeltaH1}
     \Delta H_I
   &=&\int^{t_f}_{t_i} dt\, \left(  \frac{\partial {H}_I}{\partial{\bm \alpha}}\dot {\bm \alpha}+
   \frac{\partial {H}_I}{\partial{\bm\beta}}\dot {\bm \beta}\right).
\end{eqnarray}
In what follows we choose for ${\bm\beta}$
a selection of:
\begin{eqnarray}\label{beta}
     {\bm\beta}&=&\left(c_g^2,c_m^2
    \right)
\end{eqnarray}
the gravity and matter speed of light.

\subsection{Model}\label{model}
As already stressed, 
the separation of $U$ and $L$ is part of the model, and we choose the setting  explained at the end of Section~\ref{Machset}. Thus, $I$ only takes 2 values,  $I=L,U$, with 
$L$ the visible FRW model, and $U$ an outer FRW model beyond our horizon, the two having different properties. There could be other regions (indeed there must be at least a buffer region between these two);  however these do not affect the global variables (dominated by $U$) or our local Universe, and so can be ignored. 

The symmetry reduced actions for $I=L,U$ are:
\begin{eqnarray}
    S_L&=&V_L\int dt\, [\alpha_2 \dot b a^2 +\sum_k \dot m_k T_{mk}- NH_L],\label{SL}\\
    H_L&=&a\left(-\alpha_2 (b^2+kc_g^2)+\sum_k \frac{m_k}{a^{1+3w_k}}\right),\label{HL}\\
    S_U&=&V_U\int dt\, [\alpha_2 \dot {\Bar b} \Bar a^2 +\sum_k \dot {\Bar m}_k \Bar T_{mk}- NH_U],\label{SU}\\
    H_U&=&\Bar a\left(-\alpha_2 (\Bar b^2+\Bar kc_g^2)+\sum_k \frac{{\Bar m}_k}{{\Bar a}^{1+3w_k}}\right)\label{HU}
\end{eqnarray}
(see e.g~\cite{evolution} and references therein).
Here $a$ is the expansion factor, $b$ the FRW connection variable (which in the standard theory is $\dot a$ on-shell, that is the comoving inverse Hubble length), and the fluids indexed by $k$ have been represented following~\cite{brown,gielen,gielen1,evolution}, with $w_k$ the equation of state. We use a bar for the $U$ variables and no bar for the $L$ variables. 
This defines the Poisson brackets for all variables, as well as the Hamiltonian. 

By comparing (\ref{DeltaH}) (with $I$ identified with $L$) with the fluids term in $H_L$ (cf.\eqref{HL}), we see that any past evolution in the laws of physics (or more generally any non-locality) leaves a memory that can be identified with a pressureless matter component. Indeed any integration constant in $H_L$ is exactly matched by a $w_k=0$ component. This is because a constant in the Hamiltonian is exactly identical at this level of symmetry with a pressureless component. 
This should not be confused with any pressureless fluids already present, such as baryonic or cold dark matter  (BDM or CDM). The phenomenon we are describing is not a source term for these species but, rather, it is equivalent to adding a {\it new} matter component which we will call Machian matter and label with an ingredient label $k\equiv M$.
This $k=M$ component in the fluids term of (\ref{HL}) is characterized by:
\begin{eqnarray}
      w_M&=&0\\
    m_M&=&-\int dt\, \left(  \frac{\partial {H}_L}{\partial{\bm \alpha}}\dot {\bm \alpha}+
   \frac{\partial {H}_L}{\partial{\bm\beta}}\dot {\bm \beta}\right).\label{mH}
\end{eqnarray}
There is nothing a priori forcing $m_M>0$, but since this is not a real matter component, a negative ``mass'' need not signal an instability. Furthermore, as we will see, the equivalence with dark matter only works at the FRW level. When we examine inhomogeneities and clustering properties later in this paper, some differences will be found between this effective pressureless fluid and any conventional form of dark matter, be it BDM, CDM or any other variant.

\subsection{Scenarios}\label{scenarios}
We finally need to settle on possible ``scenarios''. For the particular theories described in~\ref{ex-theories},
these follow from choices of $\bm\beta(T_{\bm\alpha})$. For the Machian setting chosen in~\ref{model}, the outer Universe $U$ then defines functions $\bm\alpha (t)$ $\bm T_{\bm\alpha}(t)$ and $\bm\beta(t)$, and these are external inputs for the $L$ dynamics. It is these functions that we wish to classify here. Although there are exceptions (e.g., power-laws, such as in~\cite{Barrow1,Barrow2}) many scenarios for varying constants are ``phase transitions''~\cite{VSL0,VSL}, presumably ``first order'', since they were modelled as step functions. 

Along these lines we introduce the classification (illustrated with parameters $\bm\beta$, but it applies to other variables):
\begin{itemize}
    \item Pulses or bleeps:
    \begin{equation}
        {\bm \beta}=\delta_{\bm \beta} \,\delta(t-t_\star)+{\bm \beta}_0,
    \end{equation}
where ${\bm \beta}_0$ is the baseline value of the variable. 
    \item Steps or jumps:
    \begin{equation}
        {\bm\beta}={\bm \beta}_-+\Delta {\bm \beta} \,H(t-t_\star).
        \end{equation}
        where the variable jumps from one constant value to another (here from ${\bm \beta}_-$ to ${\bm \beta}_+={\bm \beta}_-+\Delta {\bm \beta} $). 
    \item Cusps: a discontinuity in the derivative of a variable. 
\end{itemize}
Since these profiles are related by differentiation and integration, all variables in a given scenario will usually fall into one of these categories, or a combination thereof.






\section{A simple illustration}\label{illustration}
As a simple illustration we consider a theory where the only $\bm \alpha$ in (\ref{alpha}) is the Planck mass  $\alpha_2$ (and so the only clock is $T_R$) and the only $\bm\beta$ in (\ref{beta}) is $c_g^2$. The non-locality is therefore  encapsulated in the evolution potential $c_g^2=c_g^2(T_R)$. Furthermore, we assume that the Machian presiding Universe is spatially flat ($\bar k=0$), so that equations (\ref{GlobalApprox1}) and (\ref{GlobalApprox2}) reduce to:
\begin{eqnarray}
     \dot \alpha_2&\approx &
    - N\alpha_2  \frac{d c_g^2}{d T_R}
   \Bar{a}\Bar{k} =0\\
    \dot T_R&\approx &
  \frac{N}{\alpha_2} {\Bar{a}^3}\frac{\Bar\rho-3\Bar p}{2} . \label{TRmodel}
\end{eqnarray}
Hence, 
in this model the time-dependence in the Hamiltonian induced by  $c_g^2=c_g^2(T_R)$ does not entail corresponding variations in $\alpha_2$, the dual to the clock. This simplifies the calculations (see~\cite{evolution,BHevolution} for the technical complications to be found otherwise).

The local Hamiltonian therefore is:
\begin{eqnarray}
     H_L&=&N a\left(-\alpha_2 (b^2+kc_g^2)+\frac{m}{a}\right)
\end{eqnarray} 
where $\alpha_2$ does not vary and $c_g^2(t)$ results from composing the functions $c_g^2(T_R)$ and (\ref{TRmodel}) and can be seen as an input external to the $L$ dynamics. The Hamilton equations for the local Universe are:
\begin{eqnarray}
     \dot a
     &=& \{a,H_L\}=Nb\label{dotaG1}\\
    \dot b&=& \{b,H_L\}=-\frac{N}{2 a} (b^2+kc_g^2) 
    \label{dotbG1}\\
    \dot m&=& \{m,H_L\}=0,
\end{eqnarray}
where we have {\it not} used $H_L=0$. Thus, unlike in standard cosmology, we do not have a Friedman equation (equivalent to $H_L=0$). In addition, combining the first and second equation leads to the Raychaudhuri equation stripped of its standard simplification using the Friedman equation. The last equation is a conservation equation, valid in this case (since $\rho$ does not depend on any $\bm\alpha$ or $\bm\beta$). 
Note that because $H_L=0$ cannot be imposed we lose the ability to redefine $N$ as a function of $a$ or $b$ (the Poisson bracket would generate a non-vanishing term in the derivatives of $N$ multiplied by $H_L$ leading to new terms in these equations). In what follows we assume $N=1$, that is we work with a global lapse function which reduces to 1 in the local FRW patch.


Let us now assume a scenario with a pulse in $c_g^2$ at $t=t_\star$:
\begin{equation}
  c_g^2=c^2_{g0}+\delta_{c_g^2}\, \delta(t- t_\star), 
\end{equation}
(so that  $[\delta_{c_g^2}]=L^2/T$). 
Eqs.~(\ref{dotaG1}) and (\ref{dotbG1}) imply a cusp in $a$ (at $a(t_\star)=a_\star$) and a step in $b$ with:
\begin{equation}
    \Delta b=-\frac{k\delta_{c_g^2}}{2a_\star}.
\end{equation}
Hence, in addition to suffering a bleep (that will be forgotten) $H_L$ feels a step at $t_\star$ (that will persist) given by:
\begin{equation}\label{DelHpulse}
    \Delta H_L=- a_\star \alpha_2 \Delta b^2=
    - a_\star \alpha_2 (2b_{-} \Delta b +(\Delta b)^2)
\end{equation}
where $b_-$ is the value of $b$ just before the jump. This is consistent with 
(\ref{DeltaH1}), which gives after an integration by parts (and use of $\int dx\, \delta(x)H(x)=1/2$):
\begin{equation}
    \Delta H_L= \alpha_2 \int dt\, kbc_g^2=\alpha_2 k\left(b_-+\frac{\Delta b}{2}\right)\delta_{c_g^2}
\end{equation}
identical to (\ref{DelHpulse}). As explained before, this shift in the Hamiltonian presents problems for enforcing a Hamiltonian constrain after the bleep, assuming it was valid before. But this shift is equivalent to keeping the Hamiltonian constraint after the pulse, and introducing a new pressureless fluid with with $m_M=-\Delta H_L$. This is the simplest scenario we found for production of Machian matter. 

Another reasonably simple alternative results from a phase transition in $c_g^2$. Then, analysis of our equations shows that $a$ and $b$ must be continuous (with a cusp in $b$), and we find:
\begin{equation}
    \Delta H_L=-m_M=-a_\star k\alpha_2\Delta c_g^2.
\end{equation}
Other theories and scenarios can be investigated, but they all lead to similar results: generically a pressureless component is created by past non-localilty, according to expression (\ref{mH}).

Parenthetically we note that the simple model presented in this Section could be used to solve the flatness and homogeneity problems. Since the production of Machian matter is coupled to the local $k$ it could be used to fine tune its flatness. If we see fluctuations around a $k=0$ Universe as small open and closed Universes, this push to flatness would also homogenize the Universe (albeit leaving behind an isocurvature fluctuation involving conventional matter and Machian matter). However, this property is not generic, and is specific of the concrete theory used in this Section. 

\section{Compare and contrast}\label{contrast}
The equivalent Machian matter component is not merely an 
addition to the zero point Hamiltonian. Within the FRW approximation (homogeneity and isotropy) it gravitates as dark matter for all practical purposes, just as $\Lambda$ does but with different pressure and dilution rate. For example, the Raychaudhury equation contains it, since: 
\begin{equation}
    \ddot a= -\frac{N}{2 a} (b^2+kc_g^2)=-\frac{a}{2\alpha_2}(\rho+3p+\rho_M)
\end{equation}
where the first identity is Hamilton's equaton and the second uses the effective Hamiltonian constraint containing $m_M$. But even at the FRW level the equivalence breaks down beyond gravitation. For example, Machian matter cannot replace BDM in its role in Big Bang Nucleosynthesis. 

Nevertheless, at the FRW level the equivalence with CDM is complete, since gravitation is its only manifestation. 
CDM, however, is more than a required FRW background component: its clustering properties are also needed to explain the phenomenology. It is at this level that the effective new component shows some striking differences, providing opportunities for testability.

Unlike $\Lambda$, the new component can be inhomogeneous, and if so, the other forms of matter feel its gravity. That an inhomogeneous $m_M(x)$ attracts other matter can be probed, for example, with the spherical collapse model or the Oppenheimer-Snyder model~\cite{OS,BHevolution} (the new component contributes to the ADM mass as usual), but it can also be proved in general. Adapting Section~\ref{locglob} to the generic continuous case without any symmetry reduction, we have for the total Hamiltonian:
\begin{equation}
    H_T=V_\infty {\cal H}_T=\int_{\Sigma_t} d^3 x\, {\cal H}(x)
\end{equation}
where ${\cal H}_T$ is the total Hamiltonian density and $V_\infty=\int_{\Sigma_t}  d^3 x$ (either finite or with a limiting procedure implied). This is defined so that the $V_\infty$ factors cancel out in:
\begin{eqnarray}
    \dot\alpha&=&\{\alpha,N H_T\}=N\frac{\partial {\cal H}_T}{\partial T_\alpha}\label{ham1totb}\\
    \dot T_\alpha&=&\{T_\alpha,N H_T\}=
    -N\frac{\partial {\cal H}_T}{\partial \alpha}.\label{ham2totb}
\end{eqnarray}
(where $N=N(t)$), just as in (\ref{ham1tot}) and (\ref{ham2tot}).  
Then, the expressions at the end of Section~\ref{locglob} become in general:
\begin{eqnarray}
     m_M(x)&=&-\int dt\,\left(  \frac{\partial {\cal H}(x)}{\partial{\bm \alpha}}\dot {\bm \alpha}+
   \frac{\partial {\cal H}(x)}{\partial{\bm T}_{\bm \alpha}}\dot {\bm T}_{\bm \alpha}\right)\nn\\
   &=&-\int dt\,\left(  \frac{\partial {\cal H}(x)}{\partial{\bm \alpha}}\frac{\partial {\cal H}_T}{\partial{\bm T}_{\bm \alpha}}-
   \frac{\partial {\cal H}(x)}{\partial{\bm T}_{\bm \alpha}}\frac{\partial \bar {\cal H}}{\partial{\bm \alpha}}\right)\nn
\end{eqnarray}
with:
\begin{equation}
    \tilde {\cal H}(x)={\cal H}(x)+m_M(x)=0. 
\end{equation}
Hence, the redefined action acquires a Machian mass shift as in:
\begin{equation}
    \tilde {S}=S+S_M=
    {S}-\int dt\, N(t)\int_{\Sigma_t} d^3x\, m_M(x).
\end{equation}
Evaluating the stress energy tensor $T_{\mu\nu}^M$ of the new component (with $\tilde T_{\mu\nu}=T_{\mu\nu}+T^M_{\mu\nu}$), 
by varying with respect to the inverse metric $g^{\mu\nu}$, it is convenient to choose the ADM parameterization adapted to $\Sigma_t$. Then, since the new term only depends on $N$, the only non-vanishing component (in these coordinates) is:
\begin{equation}
    T_{00}^M=
   N^2 \frac{m_M(x)}{\sqrt{h}}.
\end{equation}
(this is true even if the shift function $N^i\neq 0$, as long as the indices are downstairs). 
Given that the normal to $\Sigma_t$ in these coordinates is $n_\mu=(N,\mathbf{0})$,  we can write:
\begin{equation}
   T^M_{\mu\nu}
    =\rho_M(x)u_\mu u_\nu
\end{equation} 
where $u_\mu=n_\nu$
and:
\begin{equation}
    \rho_M=\frac{m_M}{\sqrt{h}}.
\end{equation}
This is the stress-energy tensor of a dust fluid and variation of the full action with respect to the metric shows that $T^{\mu\nu}_M$ appears in the right hand side of the Einstein equations, so it has the usual {\it active} gravitational properties, as claimed. 

However the {\it passive} gravity is very different, as can be guessed from the fact that $m_M(x)$ does not depend on time, regardless of whatever other matter may be around to ``pull it''. The density $\rho_M$ does depend on time but only via the spatial determinant of the metric $h$, so no conservation law is expected, and indeed this is verified by calculation. Recall that for conventional dust one finds independently geodesic motion for $u^\mu$ and constancy of the dust grains rest mass, $\nabla_\mu (\rho u^\mu)=0$. Instead, here $u^\mu\equiv n^\mu$ is fixed by $\Sigma_t$ (and so signals the presence of a preferred frame), and is generically non-geodesic:
\begin{equation}
    S^\alpha =P^\alpha_{\; \; \nu} \nabla_\mu T^{\mu\nu}=u^\mu\nabla_\mu u^\alpha\neq 0
\end{equation}
(with $P_{\mu\nu}=g_{\mu\nu}+u_\mu u_\nu$ the  usual projector, and $S_\alpha u^\alpha=0$). 
Independently from this, we find that:
\begin{equation}
    S=-u_\nu \nabla_\mu T^{\mu\nu}=\nabla_\mu (\rho u^\mu)=(m_M N^i)_{,i}
\end{equation}
is generically non-zero. 
This should not be surprising. Machian matter does not correspond to any grains of dust, so there is no reason why it should follow geodesics or its ``rest mass'' be conserved.

We thus see that, once we go beyond homogeneity and isotropy, our Machian matter becomes a form of ``painted-on'' dark matter. It is encrusted in the preferred frame according to a fixed $m_M(x)$ and has zero passive gravitational mass, refusing to budge from its locus regardless of the gravity of other matter and itself. And yet, it does attract other forms of matter (including visible matter). This opens it up to  being experimentally distinguished from standard dark matter, as we will briefly discuss in our final Section. 

Note that the fact that $T^{\mu\nu}_M$ is not conserved, but enters the right hand side of the Einstein equations is not contradictory, because of the implied frame dependence and consequent collapse of the Bianchi identities. We should also stress that the non-conservation of $T^{\mu\nu}_M$ is a separate effect from the non-conservation of the other $T^{\mu\nu}$  explored in~\cite{evolution}. The latter happens while non-locality/evolution is switched on; the former perdures even after it switches off. The latter can happens in the FRW approximation; the former cannot.

\section{Discussion}\label{concs}

Dark matter is usually introduced as a putative new cosmic ingredient to explain several observations. But could what we call dark matter be a memory of past evolution in the laws of physics? In this paper we considered this possibility within the context of evolution in terms of global time variables, so that our conclusions apply more broadly to any theory with non-local interactions. Hence, we probed a more general Machian idea, where the global Universe can act directly on the local physics. We focused on the legacy effect of past global interactions: their integrated effect spoils the Hamiltonian constraint even after they switch off, but this can be rectified with a redefined Hamiltonian containing a new form of matter, which for lack of a better name we called Machian matter\footnote{Other possibilities include ``memorious matter'' in a nod to Borges, and ``mnematter'' or ``mnemomatter'' in a nod to the Greek language.}. We went on to note that Machian matter mimics many features of conventional dark matter.  

``Dark matter'' is a broad term serving several aspects of phenomenology. We found that Machian matter reproduces all of its gravitational effects at the level of an homogeneous and isotropic background. Beyond that the differences are significant, and are two-fold:
\begin{itemize}
    \item  
    Although Machian matter perfectly mimics the active gravitational pull of inhomogeneous dark matter, it fails to cluster around other forms of matter or feel its own gravity. 
    \item Machian matter remains stuck in position as defined by a fixed frame, thereby physically pegging down a preferred frame. 
\end{itemize}
Thus, Machian matter is the ultimate painted-on dark matter. Computer simulations show that, in the non-linear regime, conventional dark matter occasionally fails to stay put where it is needed for phenomenology, under the pull of its own gravity and of other matter. A case in point is the dark matter needed to explain galaxy rotation curves, and its ungainly tendency to self-collapse into a cusp, in contradiction with the required profile.  Fuzzy dark matter~\cite{fuzzy,fuzzya,fuzzy1} is an alternative proposed to skirt this drawback. We do not have this problem: Machian matter stays where we put it. Of course any suggestion that the seed for any galaxy lies in $m_M(x)$ stumbles into an initial condition issue; but is this more fine-tuned than fuzzy dark matter? The question extends to clusters, lensing, the bullet cluster, etc.

Clearly there are some positive aspects of conventional dark matter which Machian matter cannot mimic: it is never meant to be a full replacement to cold dark matter. But conversely, there are some problems with dark matter which Machian matter can solve, at the expense of a designer $m_M(x)$.  Another aspect that is different is peculiar velocities. Machian matter generates peculiar flows; yet it does not partake in them. All of this may be seen as desirable, since it makes our proposal observationally distinguishable from the more parochial forms of dark matter.





We close with some general comments on the nature of our construction. One sometimes classifies cosmological setups into dark matter/energy models vs. modified gravity theories, with some ambiguity on whether they can be made equivalent. We have at hand something that does not quite fit either description. At its roots our class of theories is best described as a modified gravity, with a dark matter reinterpretion, but as Section~\ref{contrast} showed this interpretation has its limitations. Then, of course, Machian matter breaks Lorentz invariance, something the usual dark matter does not. It pegs down a preferred frame and breaks the equivalence principle, since it has active but no passive gravitational mass. 
Once again Mach's principle inspires a construction that eventually conflicts with its best intentions.



\section{Acknowledgments}
We thank S. Koushiappas for discussion, and P. Bassani and A. Lupi for assistance with terminology.  
This work was partly supported by the STFC Consolidated Grants ST/T000791/1 and ST/X00575/1.

\end{document}